\documentclass[twocolumn]{aastex63}

\received{\today}
\revised{}
\accepted{}
\shorttitle{Fast ejecta as a potential signature for neutron stars in high-mass mergers}
\shortauthors{Elias R. Most et al.}

\newcommand{\eg}{e.g.,~}
\newcommand{\ie}{i.e.,~}
\newcommand{\cf}{cf.,~}
\newcommand{\MTOV}{M_{_{\mathrm{TOV}}}}

\usepackage{amsmath}
\usepackage{amssymb}

\begin{document}

\title{Fast ejecta as a potential way to distinguish black holes from neutron stars in
  high-mass gravitational-wave events}

\correspondingauthor{Elias R. Most} \email{emost@princeton.edu}

\author[0000-0002-0491-1210]{Elias R. Most}
\affiliation{Princeton Center for Theoretical Science, Princeton University, Princeton, NJ 08544, USA}
\affiliation{Princeton Gravity Initiative, Princeton University, Princeton, NJ 08544, USA}
\affiliation{School of Natural Sciences, Institute for Advanced Study, Princeton, NJ 08540, USA}

\author[0000-0002-6400-2553]{L. Jens Papenfort}
\affiliation{Institut f\"ur Theoretische Physik, Goethe Universit\"at,
Max-von-Laue-Str. 1, 60438 Frankfurt am Main, Germany}
\author[0000-0001-9781-0496]{Samuel Tootle}
\affiliation{Institut f\"ur Theoretische Physik, Goethe Universit\"at,
Max-von-Laue-Str. 1, 60438 Frankfurt am Main, Germany}

\author[0000-0002-1330-7103]{Luciano Rezzolla}
\affiliation{Institut f\"ur Theoretische Physik, Goethe Universit\"at,
Max-von-Laue-Str. 1, 60438 Frankfurt am Main, Germany}
\affiliation{School of Mathematics, Trinity College, Dublin 2, Ireland}
\affiliation{Frankfurt Institute for Advanced Studies, Ruth-Moufang-Str. 1, 60438 Frankfurt am Main, Germany}
\affiliation{Helmholtz Research Academy Hesse for FAIR,
Max-von-Laue-Str. 12, 60438 Frankfurt am Main, Germany}



\begin{abstract}
  High-mass gravitational-wave events in the neutron-star mass range,
  such as GW190425, have recently started to be detected by the
  LIGO/Virgo detectors. If the masses of the two binary components fall
  in the neutron-star mass range, such a system is typically classified
  as a binary neutron-star system, although the detected
  gravitational-wave signal may be too noisy to clearly establish a
  neutron-star nature of the high-mass component in the binary and rule
  out a black hole--neutron star system for such an event. We here show
  that high-mass binary neutron-star mergers with a very massive
  neutron-star primary close to the maximum-mass limit, $m_1 \gtrsim 2.2
  \, M_\odot$, produce at merger fast dynamical mass ejecta from the
  spin-up of the primary star at merger. By simulating the merger of
  black hole--neutron star systems of exactly the same masses and spins,
  we show that these fast ejecta are entirely absent, if the primary is
  instead a black hole. In addition, we find that both systems leave
  almost identical amounts of baryon mass behind, which is not
  immediately accreted by the black hole. This implies that both systems
  will likely have comparable electromagnetic afterglow emission stemming
  from the remnant disk. Hence, fast ejecta at merger have the potential
  to distinguish neutron stars from black holes in high-mass mergers,
  although these ejecta may be challenging to detect observationally.
\end{abstract}


\section{Introduction}
\label{sec:intro}

Since the first detection of a multi-messenger gravitational event
involving the merger of two neutron stars (NSs), GW170817
\citep{Abbott2017_etal}, two additional gravitational-wave events,
GW190425 \citep{Abbott2020} and GW190814 \citep{Abbott2020b}, have been
detected where at least one of the binary components lies within the NS
mass range. In the case of GW190814, which features a $23\, M_\odot$
black hole (BH) and a $2.6\, M_\odot$ companion, the nature of the
secondary has not been determined and is also at odds with our current
understanding of compact binary system formation channels
\citep{Zevin2020, Hamers2020, Safarzadeh2020b, Kinugawa2020, Liu2020,
  Lu2020}. As a result, the secondary of GW190814 could either have been
a BH (the lightest ever detected, but likely produced by the collapse of
a NS) or a stable NS (the heaviest ever detected) \citep{Most2020d,
  Tan2020, Zhang2020, Vattis2020, Fattoyev2020, Tsokaros2020,
  Godzieba2020, Essick2020, Lim2020, Roupas2020, Sedrakian2020, Tews2020,
  Biswas2020, Dexheimer2020, Cao2020, Kawaguchi2020}. Conversely,
GW190425 with a total mass of $3.4\, M_\odot$ lies firmly in the NS mass
range, \ie with individual masses being less than current constraints on
the maximum mass of nonrotating NSs, $\MTOV$, either from
gravitational-wave events
\citep{Margalit2017,Rezzolla2017,Ruiz2017,Shibata2019} or from pulsar
observations \citep{Cromartie2019,Antoniadis2013}. Since no
electromagnetic counterpart has been observed for either event
\citep{Coughlin2020,Vieira2020,Ackley2020,Watson2020,Dobie2020,Andreoni2020}
and the tidal deformability constraints remain inconclusive
\citep{Abbott2020b}, a BH--NS as the source of this event cannot be fully
ruled out \citep{Foley2020,Kyutoku2020}. This raises the general question
of how the merger of two NSs can be distinguished from the merger of a NS
and a BH when the gravitational-wave detection is not sufficiently
accurate to place a lower bound on the tidal deformability of the system.

One potential avenue is to check for consistency of the masses with the
allowed mass range for NSs \citep{Foley2020,Most2020c}, which is
complicated by the degeneracy between the mass ratio and the effective
spin in the gravitational-wave phase \citep{Harry2018}. While the same
arguments have been applied also to GW190814
\citep{Most2020d,Zhang2020,Biswas2020}, they have largely remained
inconclusive and can provide deeper insights only if it was known a
priori if the system contained either a NS or a BH primary.

Another possibility to constrain the nature of the system is through the
use of optical counterparts, in particular through the kilonova afterglow
\citep{Metzger2017}. Numerical investigations have shown that BH--NS
systems typically feature mass outflows that are different from those of
NS--NS systems
\citep[\eg][]{Foucart2012,Foucart2013a,Kyutoku2015,Kyutoku2018}. The
dynamical mass ejection in a BH--NS system will only stem from the
unbound parts of a tidally disrupted NS
\citep{Rosswog05,Kyutoku2013,Kyutoku2015}, whereas NS--NS mergers will
also feature a collision shock and mass ejection from the stable remnant
\citep{Baiotti08,Foucart2015,Sekiguchi2015,Lehner2016,
  Bovard2017}. Furthermore, the presence of a blue kilonova associated
with shock heated ejecta coming from the surface of the remnant NS
\citep{Perego2016,Fujibayashi2017}, is different to the predominantly red
kilonova emission associated with the long-term mass ejection from the
accretion disk
\citep{Fernandez2015,Fernandez2015b,Siegel2017,Miller2019b}.  Consistent
with astrophysical expectations from population synthesis models, most
numerical studies of BH--NS mergers have focussed on intermediate mass
ratios $q^{-1} := m_1/m_2 = 4-8$, where $m_{1,2}$ are the masses of the
two binary components,
\citep{Etienne:2008re,Kyutoku2011,Kyutoku2015,Foucart2011,Foucart2013a},
with BHs outside of the anticipated mass gap \citep{Kruckow2018}. Using
parameter ranges consistent with GW170817, \citet{Hinderer2018} have
performed the first direct comparison of a NS--NS merger with a BH--NS
merger. In these simulations they contrasted the evolution of a stable
remnant NS with that of the accretion disk in the BH--NS case, in
addition to tidal deformability constraints from the gravitational
waveform. While they find that a BH--NS nature is difficult to reconcile
with the EM counterparts of GW170817, it cannot be fully ruled out.
\citet{Kyutoku2020} have focussed on the observational implications of a
BH--NS in the GW190425 mass range, whereas \citet{Foucart2019,Hayashi2020} have
performed general investigations of near equal-mass NS--NS mergers,
studying in particular the mass ejection and remnant disk masses
\citep{Foucart2018b}.

One of the open problems in these studies remains the formation history
of the low-mass BH in these binary systems. In fact, while forming BHs
with $<5\, M_\odot$ remains very challenging \citep{Zevin2020,Ertl2020}, forming a
BH with masses $m_{\rm BH} < \MTOV$ remains even more challenging and
typically involves invoking primordial BH formation
\citep{GarciaBellido96}. Yet, recent observations provide a basis for
assuming the formation of very heavy NSs \citep{Cromartie2019} beyond the
currently assumed galactic pulsar-mass distribution
\citep{Alsing2017,Tauris2017}. Although overall less likely, if such a
star had a massive companion, accretion induced collapse
\citep{Qin98,Rueda2012,Fryer2014}, as suggested for Cyg X-3
\citep{Zdziarski2013} or GRB090618 \citep{Izzi2012}, or removal of
angular momentum, \eg via pulsar spin-down \citep{Falcke2013}, of a very
massive rotationally supported NS could lead to the formation of a BH
with masses $m_{\rm BH} \approx \MTOV$. Such systems are very interesting
in the present context, because they do not only lie within the mass
range of GW190425-like binaries, but they also represent a region of the
mass range where NS--NS and BH--NS could potentially coexist. Making use
of quasi-universal relations relating the mass ratio and the effective
spin
\begin{equation}
  \label{eqn:chitilde}
  \tilde{\chi}
  :=\frac{m_1 \chi_1 + m_2 \chi_2 }{m_2 + m_1} = \frac{\chi_1}{1+q}\left( 1
  + q\frac{\chi_2}{\chi_1} \right)\,,
\end{equation}
it is possible to perform targeted simulations aimed at identifying
potential differences between BH--NS and NS--NS merger for GW190425-like
events. We here present a comparison between BH--NS and
NS--NS mergers in the relevant parameter range, \ie for mass ratios
$q^{-1} \approx 2$ and primary spins $0 \leq \chi_1 \leq 0.53$.

This paper is structured as follows, in Sec. \ref{sec:methods} we
summarize the numerical methods used in this study. Section
\ref{sec:results} presents a detailed comparison of the dynamical mass
ejection of BH--NS and NS--NS mergers, and identifies a fast ejection
signatures characteristic for NS--NS merger in this mass range. Finally,
Sec. \ref{sec:discussion} concludes with a discussion of the impact of
these results.

\section{Methods}
\label{sec:methods}

To model the evolution of the compact binaries in full general
relativity, we numerically solve the Einstein field equations using a
constraint-damping formulation of these equations
\citep{Bernuzzi:2009ex,Alic:2011a} [in these specific simulations we have
  used the Z4c formulation of \citet{Hilditch2013}], coupled to the
equations of general-relativistic magnetohydrodynamics (GRMHD)
\citep{Duez05MHD0,Shibata05b,Giacomazzo:2007ti} using the
\texttt{Frankfurt\-/IllinoisGRMHD} code (\texttt{FIL})
\citep{Most2019b,Most2018b}, which is derived from the
\texttt{IllinoisGRMHD} code \citep{Etienne2015}, but implements
high-order (fourth) conservative finite-difference methods
\citep{DelZanna2007} and can handle temperature and electron-fraction
dependent equations of state (EOSs). Neutrino cooling and weak
interactions are included in the form of a neutrino leakage scheme
\citep{Ruffert96b,Rosswog:2003b,Galeazzi2013}.

\texttt{FIL} makes use of the Einstein Toolkit infrastructure
\citep{loeffler_2011_et}, and, in particular, of the adaptive
mesh-refinement driver \texttt{Carpet} \citep{Schnetter-etal-03b}, which
allows to decompose the numerical domain in a box-in-box approach. The
simulations presented here make use of nine nested Cartesian boxes and a
refinement ratio of two. The outermost box stretches to $\simeq 6000\,\rm
km$ in each direction and the innermost box covering the BH/primary NS and
(initially) the NS extends to about $17.7\, \rm km$ with a finest
resolution of $\simeq 215\, \rm m$. While we have included a buried
dipolar magnetic field of $\simeq 10^{14}\, \rm G$ in the secondary.  For
consistency with the BH--NS data set, where the primary is a BH, we have
not included a magnetic field in the primary NS of the NS--NS
system.  We find that their presence does not influence the dynamics and
mass ejection at merger.  Rather, magnetic fields will affect the
post-merger evolution of the remnant accretion disk \citep[see,
  \eg][]{Giacomazzo:2010,Dionysopoulou2015}, which we will present in a
separate study.

We model the initial binary configurations via the solution of the
Einstein constraint equations for either BH--NS
\citep{Taniguchi05,Grandclement06,Taniguchi07}, or NS--NS
\citep{Tichy12,Tacik15,Tsokaros2015,Papenfort2020} binaries. More
specifically, the initial data is constructed in the extended conformal
thin-sandwich (XCTS) formalism \citep{Pfeiffer:2002iy, Pfeiffer:2005}. In
the case of BH--NS binaries, the BH is modelled using excision boundary
conditions \citep{Grandclement06}, imposing the spin via a shift
condition on the horizon \citep{Caudill:2006hw}. The secondary NS, on the
other hand, is assumed to be irrotational. The initial data is
constructed using the publicly available \texttt{LORENE} code
\citep{Grandclement06}. On import onto the Cartesian domain, the excised
interior of the BH is filled using simple eight-order extrapolation
\citep{Etienne2007a,Etienne:2008re}. In the case of NS--NS binaries,
instead, the spinning NS is constructed following the formalism proposed
by \cite{Tichy12}, for which the velocity field of the star is a
combination of an irrotational and of a uniformly rotating part. The
initial data is then computed using the \texttt{Kadath} spectral library
\citep{Grandclement09}, which has recently been extended for compact
binary initial data \citep{Papenfort2020}.  In this way, we are able to
generate NS--NS initial data involving supramassive NSs with masses
exceeding $M_{_{\rm TOV}}$ and supported by the additional angular
momentum.  We model the microphysical composition of the NS and
finite-temperature effects in the merger using an EOS with realistic
nuclear forces \citep{Togashi2017}, hereafter referred to as
\texttt{TNTYST}. We also perform simulations using an EOS containing
hyperons \citep{Banik2014}, which is based on the DD2 EOS
\citep{Hempel2010}, and referred to as \texttt{BHB}$\Lambda\Phi$. With a
maximum mass of $M_{_{\rm TOV}} = 2.22\ (2.10)\, M_\odot$ and a radius of
$R_{1.4} = 11.54\ (13.22)\, \rm km$ for a $1.4\,M_\odot$ NS for the
\texttt{TNTYST} (\texttt{BHB}$\Lambda\Phi$) EOS, we can cover a range of
maximum masses \citep{Margalit2017,Rezzolla2017,Ruiz2017,Shibata2019} and
radii \citep{Annala2018,Most2018,De2018,Abbott2018b} that are consistent
with GW170817 \citep{Abbott2017_etal,Abbott2017b}.

\begin{deluxetable*}{|l|c|c|c|c|c|c|c|c|}
\centerwidetable
	\tablewidth{0pt} \tablecaption{Summary of the properties of the
          initial binaries. The columns list: the dimensionless spin
          $\chi_{\rm 1}$ of the primary, the effective spin $\tilde{\chi}
          = \chi_{\rm 1}/ (1+q)$, the mass ratio $q=m_{ 2}/m_{1}$ and the
          individual (baryon) masses of the primary, $m_{(b),1}$, and of
          the secondary, $m_{(b),2}$. All binaries have a total mass
          $M_{_{\rm ADM}}=3.6\, M_\odot$ and are at an initial separation
          of $45\,{\rm km}$. The secondary is always non-spinning,
          $\chi_2=0$.
    \label{tab:initial}}
	
	\tablehead{
    &$m_{1}\left[M_\odot\right]$  & $m_{2}\left[M_\odot\right]$ &
    $m_{b,1} \left[M_\odot\right]$ & $m_{b,2} \left[M_\odot\right]$ &$q$ & $\chi_{\rm 1}$ & $\tilde{\chi}$
    & ${\rm EOS}$
	}

	\startdata
    $\texttt{TNT.BH.chit.0.00}$ & $2.20$ & $1.40$ & -- & $1.55$ &$0.636$ & $0.00$ & $0.00$  & {\texttt{TNTYST}} \\
    $\texttt{TNT.BH.chit.0.15}$ & $2.24$ & $1.36$ & -- & $1.50$ &$0.608$ & $0.24$ & $0.15$  & {\texttt{TNTYST}} \\
    $\texttt{TNT.BH.chit.0.35}$ & $2.42$ & $1.18$ & -- & $1.28$ &$0.486$ & $0.52$ & $0.35$  & {\texttt{TNTYST}} \\
	\tableline                                                                                
    $\texttt{TNT.NS.chit.0.00}$ & $2.20$ & $1.40$ & $2.66$ & $1.55$ &$0.636$ & $0.00$ & $0.00$  & {\texttt{TNTYST}} \\
    $\texttt{TNT.NS.chit.0.15}$ & $2.24$ & $1.36$ & $2.70$ & $1.50$ &$0.608$ & $0.24$ & $0.15$  & {\texttt{TNTYST}} \\
    $\texttt{TNT.NS.chit.0.35}$ & $2.42$ & $1.18$ & $2.87$ & $1.28$ &$0.486$ & $0.52$ & $0.35$  & {\texttt{TNTYST}} \\
    	\tableline
	\tableline                                                                                               
    $\texttt{BHBLP.BH.chit.0.00}$ & $2.10$  & $1.50$ & -- & $1.65$ &$0.636$ & $0.00$ & $0.00$  & {\texttt{BHB}$\Lambda\Phi$} \\
    $\texttt{BHBLP.BH.chit.0.15}$ & $2.14$  & $1.46$ & -- & $1.60$ &$0.608$ & $0.24$ & $0.15$  & {\texttt{BHB}$\Lambda\Phi$} \\
	\tableline                                                                                  
    $\texttt{BHBLP.NS.chit.0.00}$ & $2.20$ & $1.40$ & $2.40$ & $1.65$ &$0.636$ & $0.00$ & $0.00$  & {\texttt{BHB}$\Lambda\Phi$} \\
    $\texttt{BHBLP.NS.chit.0.15}$ & $2.24$ & $1.36$ & $2.48$ & $1.60$ &$0.608$ & $0.24$ & $0.15$  & {\texttt{BHB}$\Lambda\Phi$} \\
	\enddata

\end{deluxetable*}
\begin{figure}[t]
  \centering
  \includegraphics[width=0.48\textwidth]{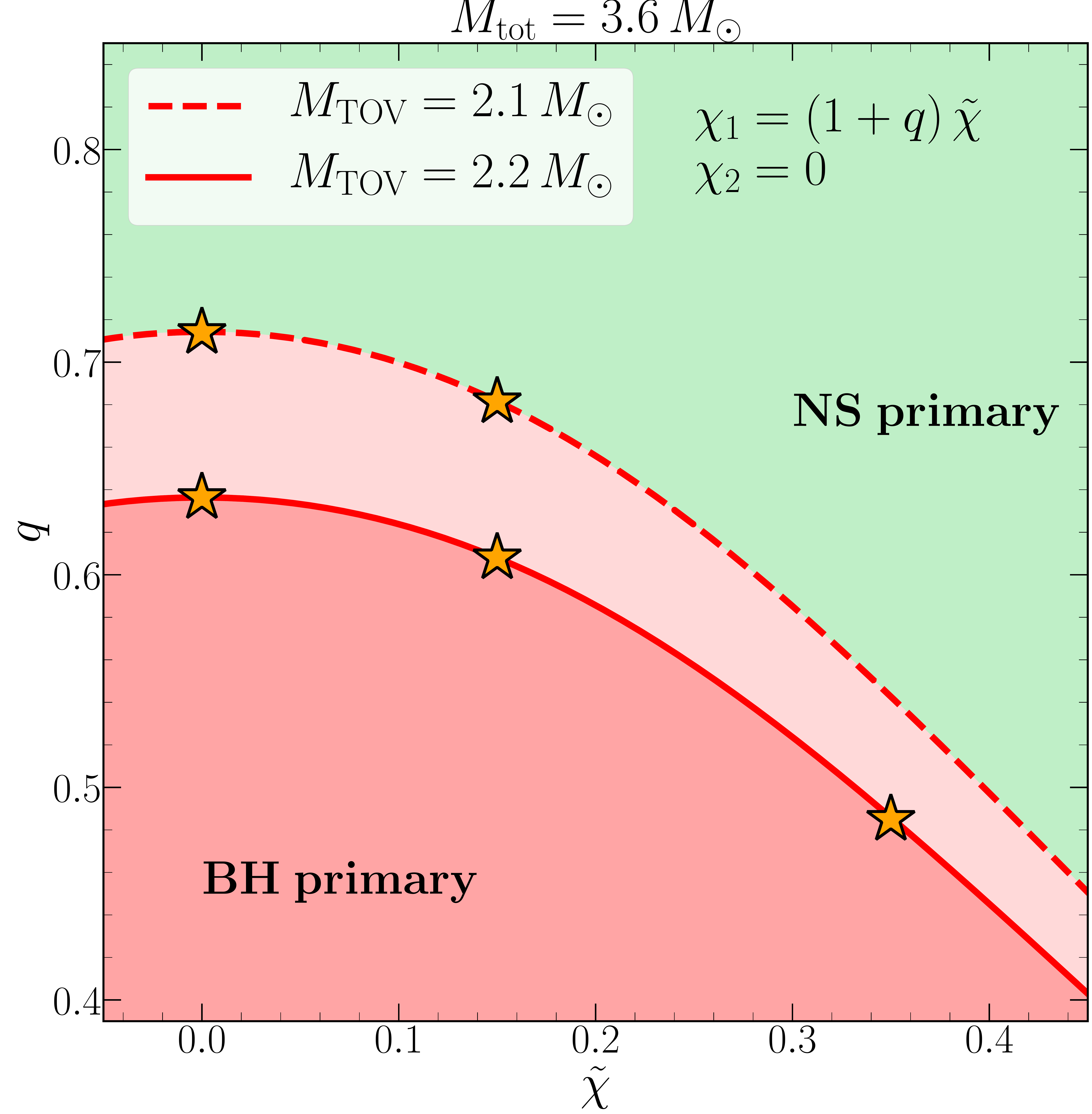}
  \caption{Schematic overview of the allowed area (green) and the
    excluded area (red) for the masses in NS--NS systems. The solid and
    dashed red lines are computed for a spin distribution, $\chi_1 =
    \left( 1+q \right) \tilde{\chi}$ and $\chi_2=0$, in terms of the
    effective spin $\tilde{\chi}$, mass ratio $q$ and total mass $M_{\rm
      tot}$. Assuming $M_{\rm BH} \simeq M_{_{\rm TOV}}$, the red lines
    outline a region for which is potentially shared by NS--NS and BH--NS
    binaries.}
  \label{fig:overview}
\end{figure}

\begin{figure*}[t]
  \centering
  \includegraphics[width=\textwidth]{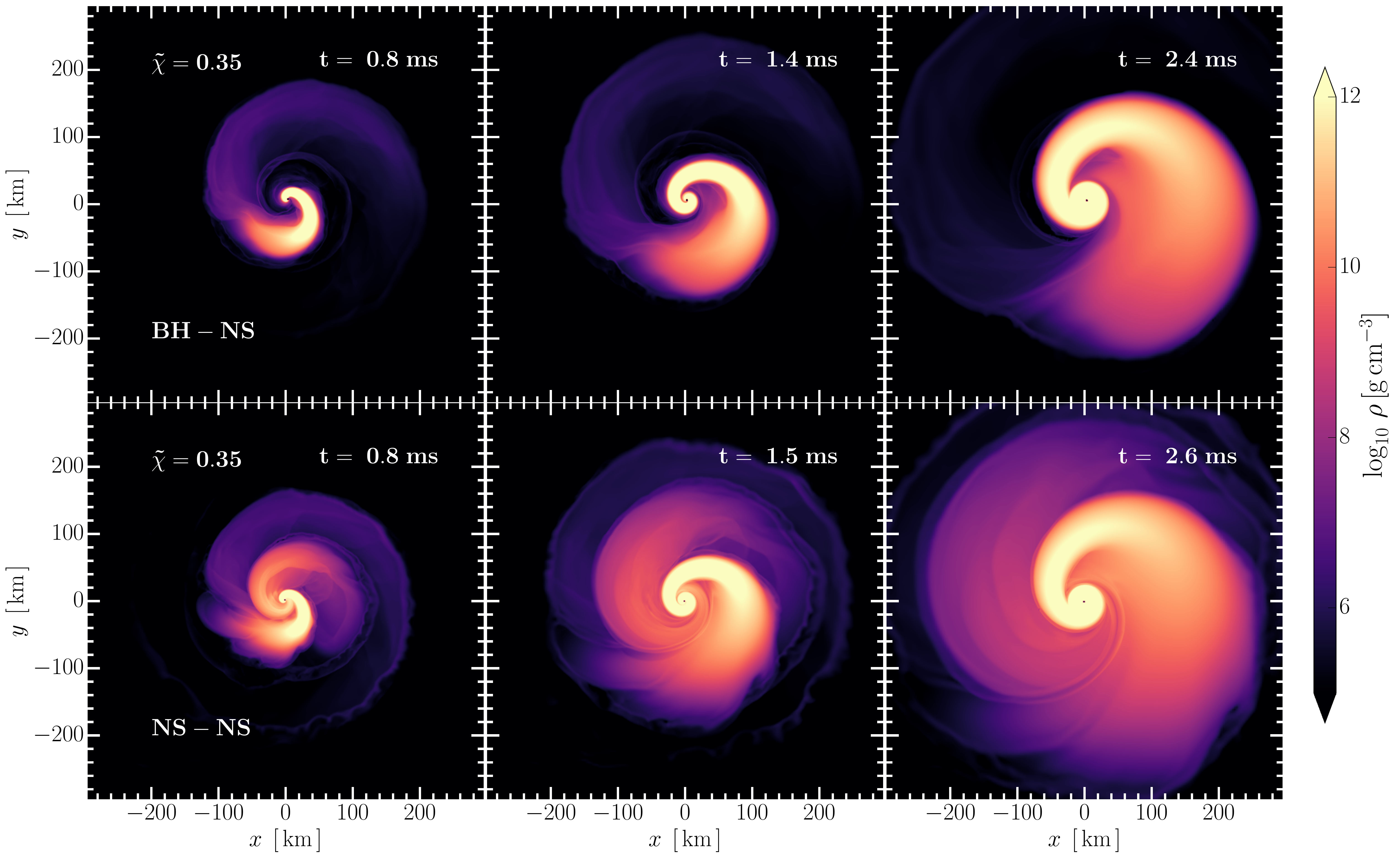}
  \caption{Tidal disruption for high-mass NS--NS (bottom) and BH--NS
    mergers (top) for effective spin $\tilde{\chi}=0.35$, computed using
    the \rm{TNTYST} EOS. Shown are three different times at merger.
    The NS--NS merger clearly features a second mass-ejection arm. }
  \label{fig:XY}
\end{figure*}

Our attention here is particularly focused on the potential differences
between NS--NS and BH--NS mergers in the near equal-mass
regime. Inspired by the detection of GW190425 with a total mass
$M=3.4^{+0.3}_{-0.1}\, M_\odot$, we adopt a total mass of $M= 3.6\,
M_\odot$ for our binary systems. This guarantees that even when
considering small mass ratios of $q^{-1} \approx 2$, the secondary mass
$m_2 = q M / \left( 1+q \right) > 1.17\, M_\odot$, which is consistent
with the proposed lower bounds on the NS mass \citep{Suwa2018}. Assuming
that no BHs are likely to be formed with masses $M_{\rm BH} < M_{_{\rm
    TOV}}$, we focus on the limiting value of $M_{\rm BH} \simeq M_{_{\rm
    TOV}}$, which would model a BH--NS system derived from NS--NS binary
where the BH is produced by means of a continued mass accretion
\citep{Safarzadeh2020b}. NSs in this mass range either have to be
nonrotating and have masses below $M_{_{\rm TOV}}$, or be rapidly
rotating if they have larger masses but still $\lesssim 1.2\, M_{_{\rm
    TOV}}$ \citep{Breu2016}. For fixed $M_{_{\rm TOV}}$, the threshold
configuration for such a NS--NS system can be computed using
quasi-universal relations \citep{Most2020c}. Following \citet{Most2020c},
we adopt three fiducial effective spin parameters $M \tilde{\chi} = m_1
\chi_1 + m_2 \chi_2$, where $m_i$ and $\chi_i$ are the individual spins
of the two binary components. We further assume that the primary reached
its high mass by accretion from the progenitor star of the secondary,
and, thus, also received additional angular momentum
\citep{Tauris2017}. For this reason, we find it reasonable to assume that
the primary spins much faster than the secondary, \ie
\begin{align}
\chi_1 \gg \chi_2 \approx 0\,,
\end{align}
which leads to [\cf Eq. \eqref{eqn:chitilde}]
\begin{align}
  \tilde{\chi} \approx \chi_1/\left(1 +q\right)
  \,.
\end{align}
We then choose $\tilde{\chi} = \left[ 0.00, 0.15, 0.35\right]$, which
corresponds to primary masses $m_1 = \left[2.20, 2.24, 2.42\right]\,
M_\odot$, mass ratios $q = \left[ 0.636, 0.608,0.486 \right]$ and primary
spins $\chi_1=\left[ 0.00, 0.24, 0.52 \right]$ for the \texttt{TNTYST}
EOS. For the \texttt{BHB}$\Lambda\Phi$ EOS we instead use $\tilde{\chi} =
\left[ 0.00, 0.15\right]$ \footnote{We have not included models with spins
  $\tilde{\chi} = 0.35$ for the \texttt{BHB}$\Lambda\Phi$ EOS, since the
initial data could would always converge to a solution for which the
primary lies on the dynamically unstable branch. This issue will be
addressed in a future release of the initial data code
\citep{Papenfort2020}.}, which corresponds to primary masses $m_1 =
\left[2.1, 2.14\right]\, M_\odot$, mass ratios $q = \left[ 0.714, 0.681
  \right]$, and primary spins $\chi_1=\left[ 0.00, 0.25 \right]$.
The properties of the simulated components in the binaries are collected in
Table \ref{tab:initial} and their position in the $(q,\tilde{\chi})$
plane is marked with stars in Fig. \ref{fig:overview}.

\begin{figure*}
  \centering
  \centering
  \includegraphics[width=0.49\textwidth]{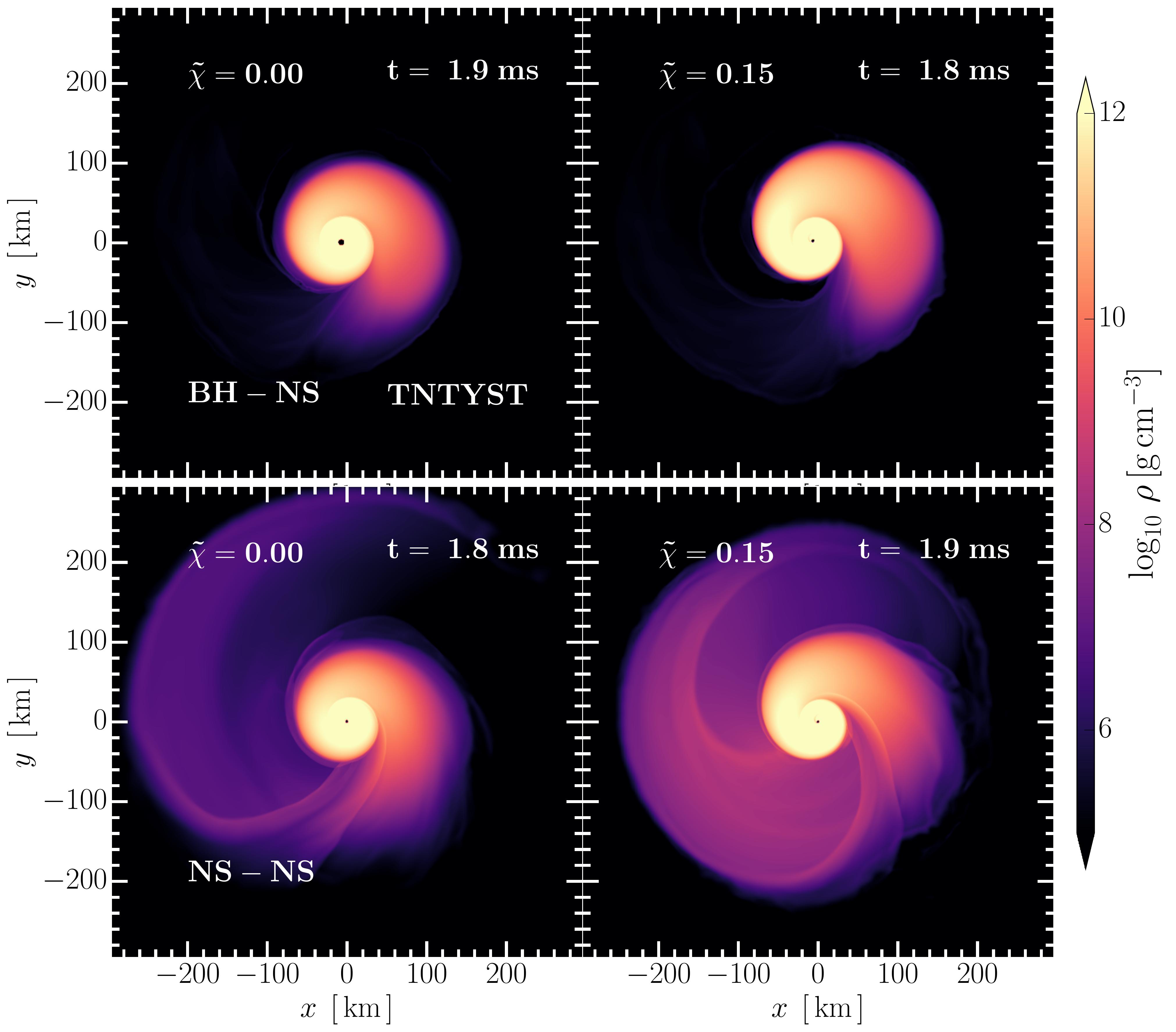}
  \includegraphics[width=0.49\textwidth]{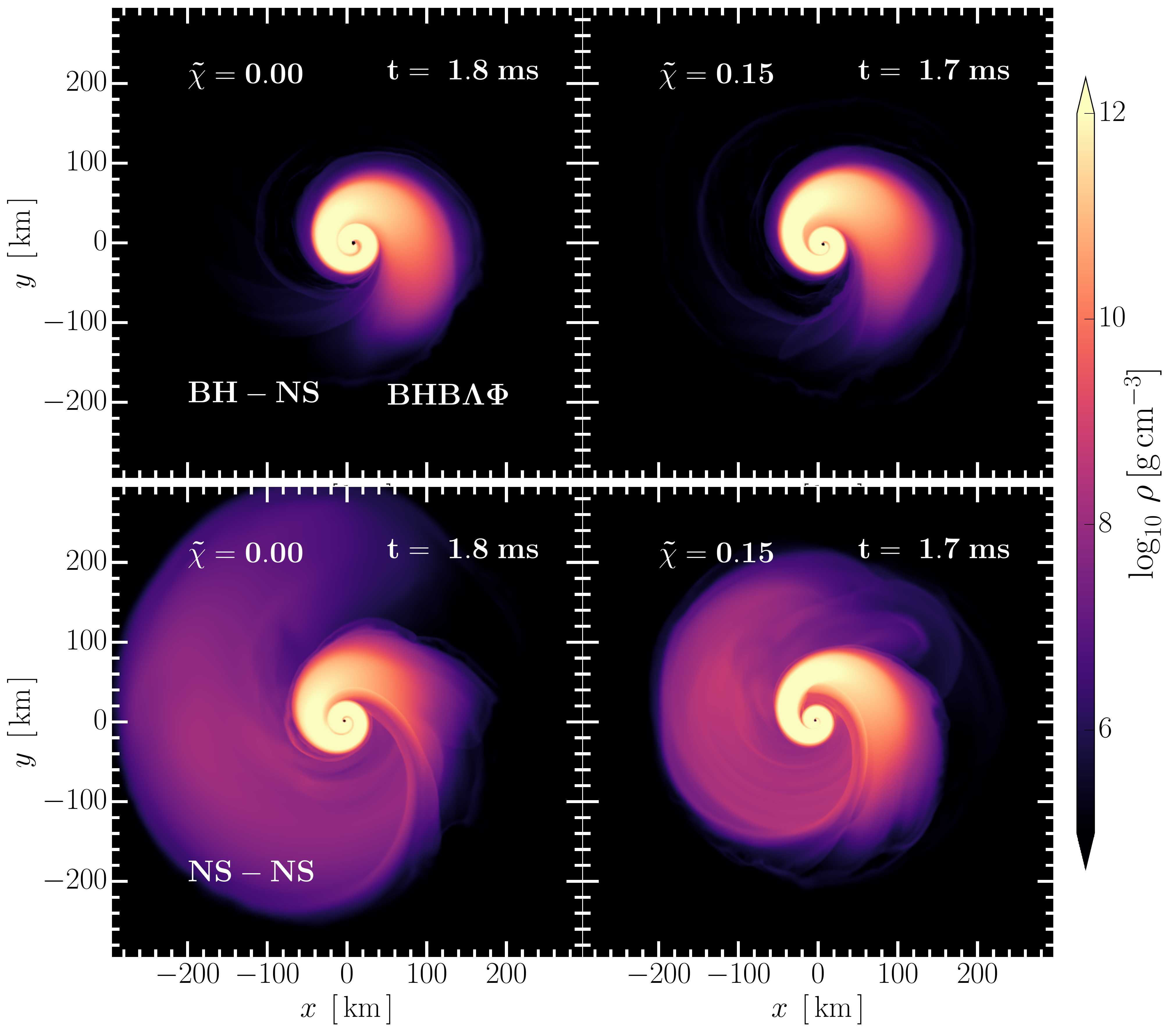}
  \caption{
	Same as Fig. \ref{fig:XY} but for $\tilde{\chi}
    = 0.00$ and for $\tilde{\chi}=0.15$.
    \textit{Left:}  Simulations using the
    \texttt{TNTYST} EOS. 
    \textit{Right:} Simulations using the 
    using the \texttt{BHB$\Lambda\Phi$} EOS.}
  \label{fig:XYB}
\end{figure*}

\section{Results}
\label{sec:results}

In the following we present the results of our simulations. In
particular, we compare the evolution of NS--NS systems to that of BH--NS
systems with the exact same mass configuration, concentrating on the
cases in which the mass of the primary is very close either to the
maximum mass of nonrotating configurations (\ie $m_1 \simeq M_{_{\rm
    TOV}}$) or to the maximum mass of rotating configurations (\ie $m_1
\simeq M_{\max}\left( \chi_1 \right)$).

\subsection{Matter dynamics}

We present the dynamics at merger in Fig. \ref{fig:XY}, which shows the
baryonic rest-mass density $\rho$ on the $(x,y)$ plane for the high-spin
system, $\tilde{\chi} = 0.35$, computed using the \texttt{TNTYST} EOS,
where the top row shows the BH--NS merger and the bottom row the
corresponding NS--NS merger at different times. The three images
illustrate the behaviour within a time interval of $2\, \rm ms$ during
the merger. Note that for the BH--NS system (top row), the NS is tidally
disrupted outside of the innermost stable circular orbit (ISCO) and the
matter and can thus form an accretion disk around the BH
\citep{ShibataTaniguchilrr-2011-6}. The long-term evolution of this disk
is then able to account for a predominantly red kilonova signal
\citep{Fernandez2015b,Siegel2017}. Contrasting this behaviour with the
evolution of a NS--NS system (bottom row), it is possible to note that
the overall picture is remarkably similar (the snapshots are chosen so as
to underline the analogies). Indeed, as first pointed out by
\citet{Dietrich2017}, the merger of two NSs at such high-mass ratios
$q^{-1} \simeq 2$, results in an evolution similar to that of BH--NS
systems. Despite the qualitative analogies between the BH--NS and NS--NS
binaries in our simulations, we crucially, find that for NS--NS systems
an additional arm of mass ejection rapidly emanating from the merger site
(see also Fig. \ref{fig:Mej}).  By carefully studying the temporal
evolution of the merger process (compare left, middle and right panels in
the bottom row), we found that this additional mass ejection represents a
second tidal tail, propagating in a direction that is opposite to that of
the main tidal tail. This is very apparent when contrasting the two
rightmost panels of Fig. \ref{fig:XY}, where it is clear that there is
essentially no mass ejected in the negative $x$-direction in the case of
the BH--NS binary (top right), while there is a large tidal tail in the
case of the NS--NS binary (bottom right). Indeed, the left-moving tidal
tail is comparable, albeit smaller, than the right-moving tail.

This different dynamical behaviour is due to the fact that in the NS--NS
binary, the largest tidal tail is not produced by the secondary (as it
necessarily is for the BH--NS binary) but likely by the spin-up of the primary.
Indeed, the tidal disruption of the secondary, when it still retains a
large fraction of its angular momentum, torques and rapidly spins-up the
primary, resulting in an extended tidal tail, which, unlike what happens
in the equal-mass case, transfers angular momentum outwards very
efficiently. This behaviour was first noted by \citet{Rezzolla:2010}
in the dynamics of unequal-mass binaries described by simple EOSs (see
also Sec. 12.5.3 of \citealt{Rezzolla_book:2013}) and explains why it is
naturally absent in BH--NS mergers: such systems contain only one NS,
which can only account for one arm of mass ejection. In contrast, NS--NS
systems will naturally have two tidal arms, which will become
increasingly more asymmetric as the mass ratio is decreased. Furthermore,
the most massive of the two tidal arms is the one generated by the
disruption of the lightest star in the binary system. In the limit of large mass
asymmetries, \eg $q^{-1} \rightarrow 2$, the heavier of the tidal arm
becomes more and more massive until it resembles the tidal arm in a
BH--NS merger. At the same time, the lighter tidal arm becomes less
massive, but does not disappear even in the limit of the maximum-mass
ratio allowed by the maximum mass of the EOS.

In summary, the formation of a double tidal-arm structure is a unique
feature of NS--NS mergers, distinguishing them from corresponding BH--NS
mergers with comparable mass, and is intrinsically associated with the
presence of a second NS in the system. As we will discuss in detail
below, the low-mass tidal arm from the secondary is composed of matter
that is ejected at substantially large velocities, \ie it can be
considered as being mostly made of ``fast ejecta'' and, for this reason,
will have a qualitatively different impact on the electromagnetic
counterpart produced in the NS--NS binaries. Note also that because of
the large initial mass of the system and the fact that the primary is
already close to the critical stability limit to collapse, the merged
object almost instantaneously collapses to a BH at merger. 
In contrast to the case of equal-mass binaries, where the BH properties do
not change significantly after its formation, variations in the mass and
spin of the BH as large as 5\% are possible for unequal-mass systems during
the subsequent evolution \citep{Rezzolla:2010}.

\begin{figure*}
\begin{center}
  \centering
  \includegraphics[width=0.35\textwidth, trim=0 0 60 0, clip]{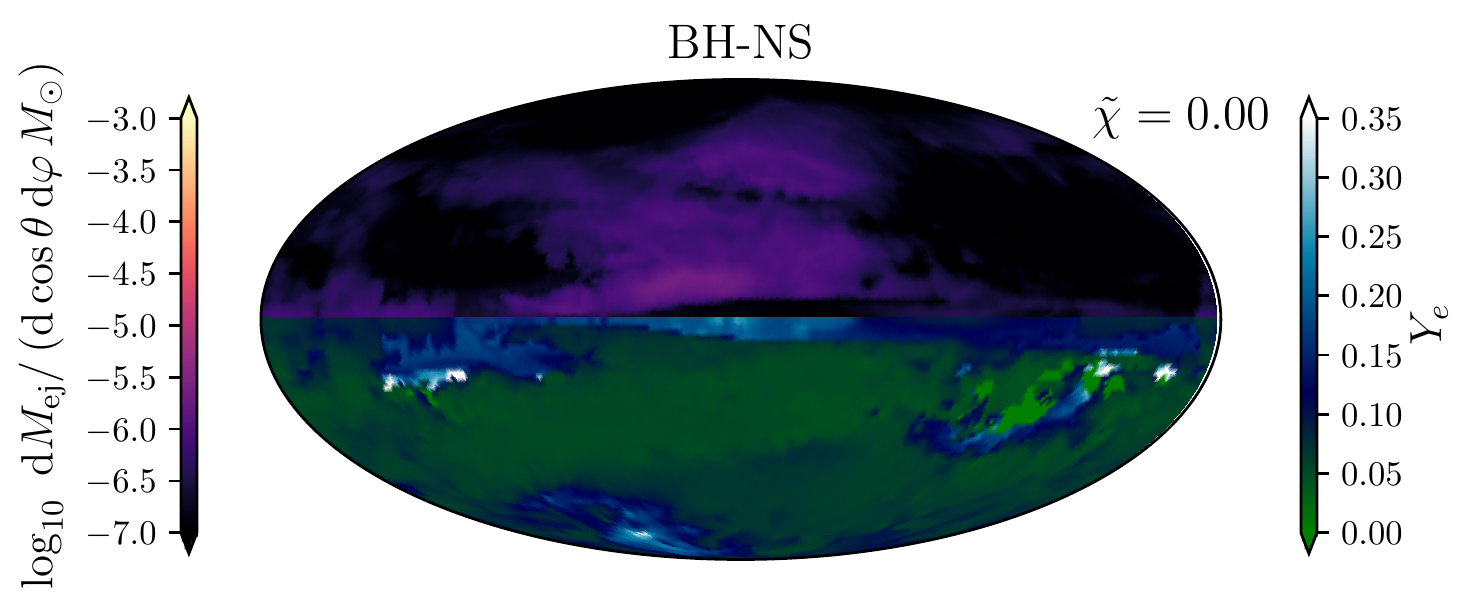}
  \includegraphics[width=0.35\textwidth, trim=60 0 0 0, clip]{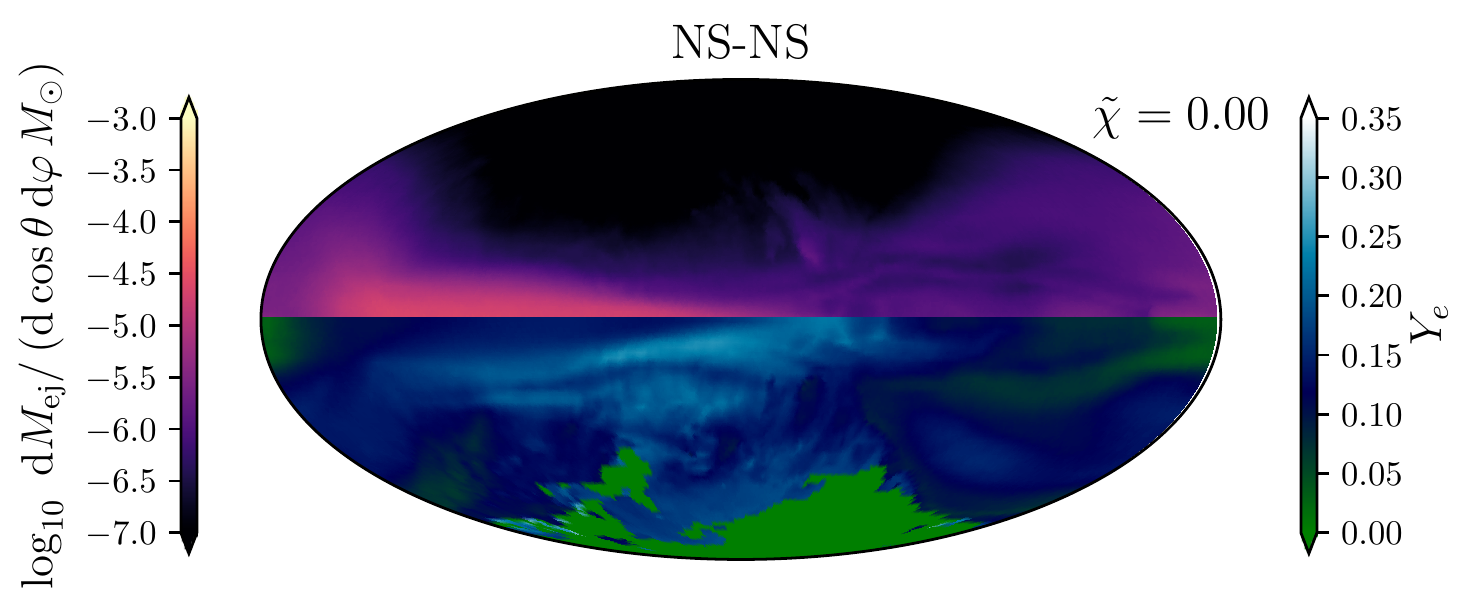}
  \\
  \includegraphics[width=0.35\textwidth, trim=0 0 60 19, clip]{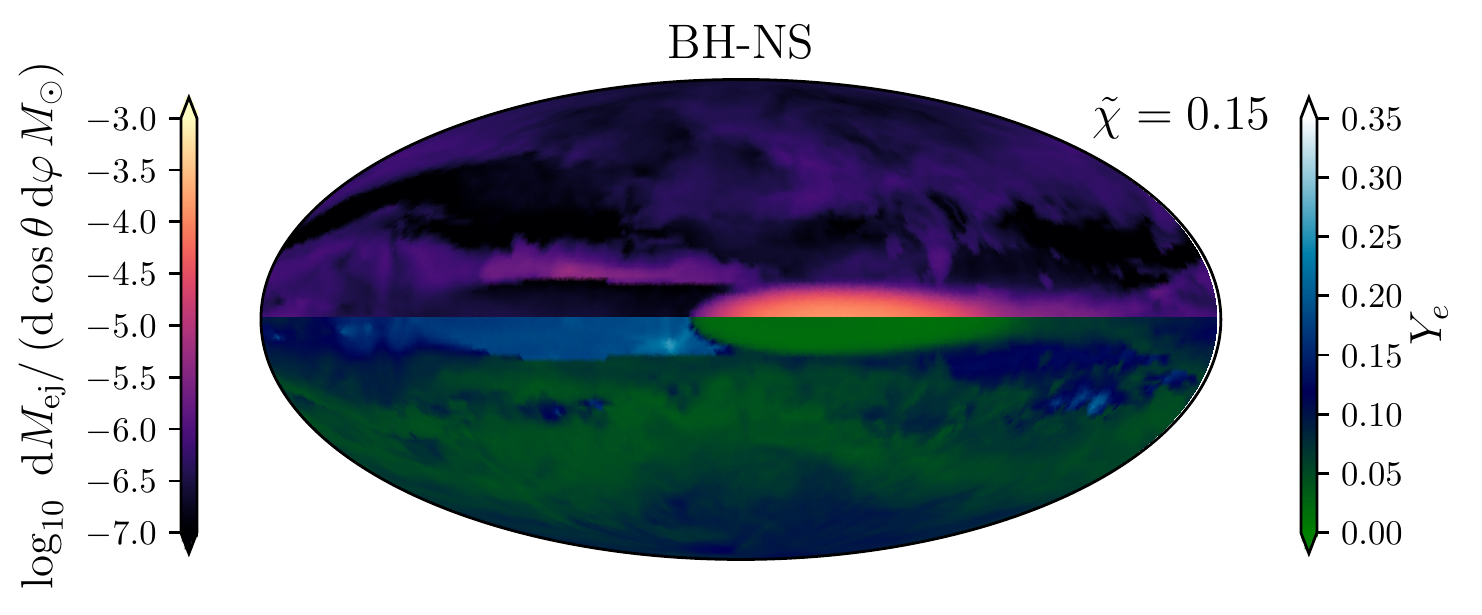}
  \includegraphics[width=0.35\textwidth, trim=60 0 0 19, clip]{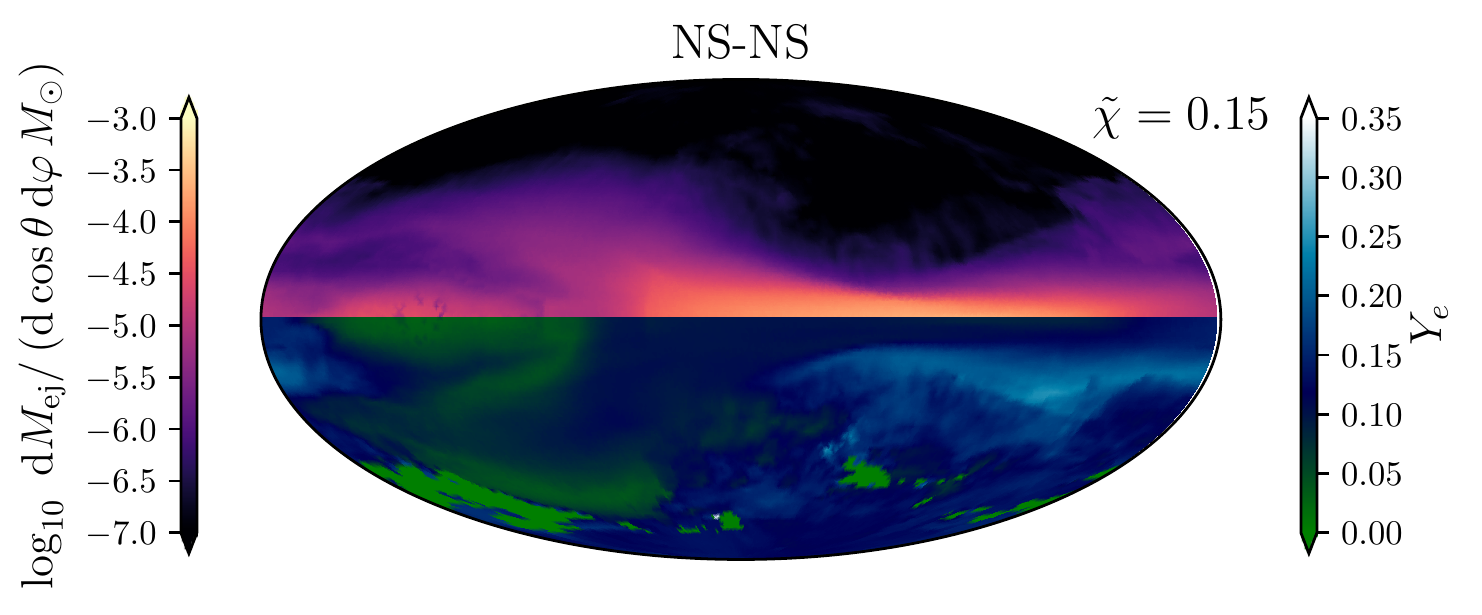}
  \\
  \includegraphics[width=0.35\textwidth, trim=0 0 60 19, clip]{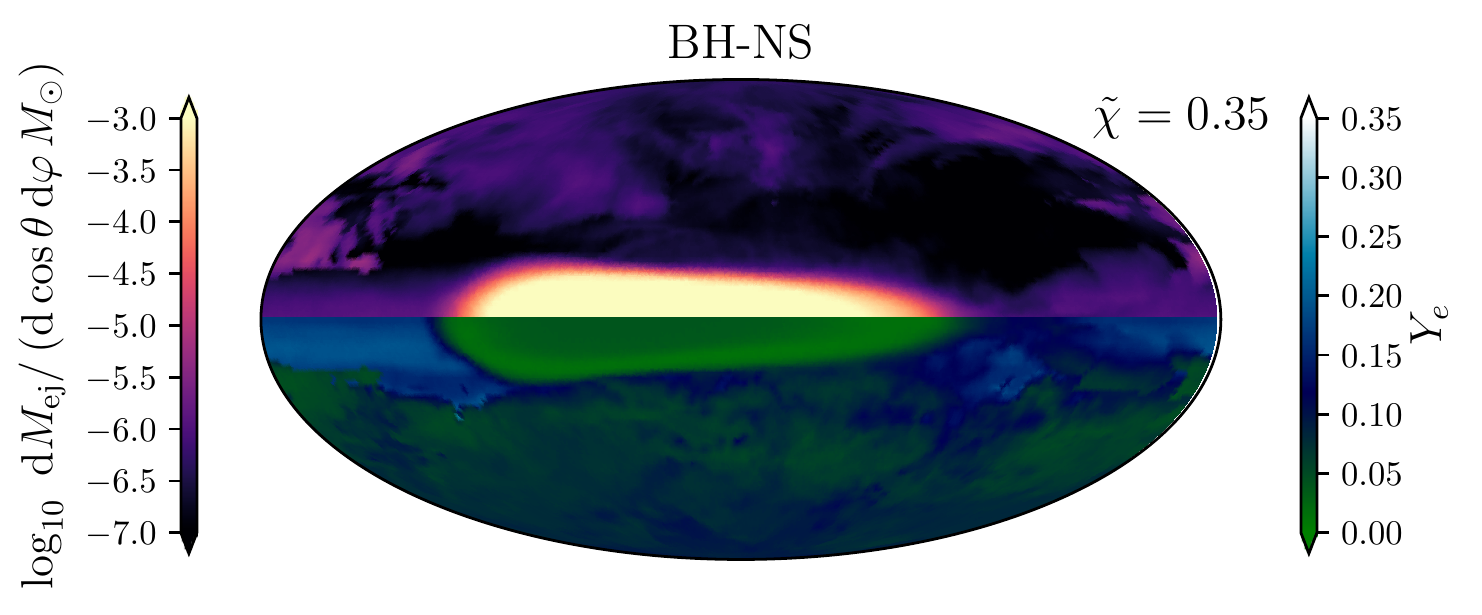}
  \includegraphics[width=0.35\textwidth, trim=60 0 0 19, clip]{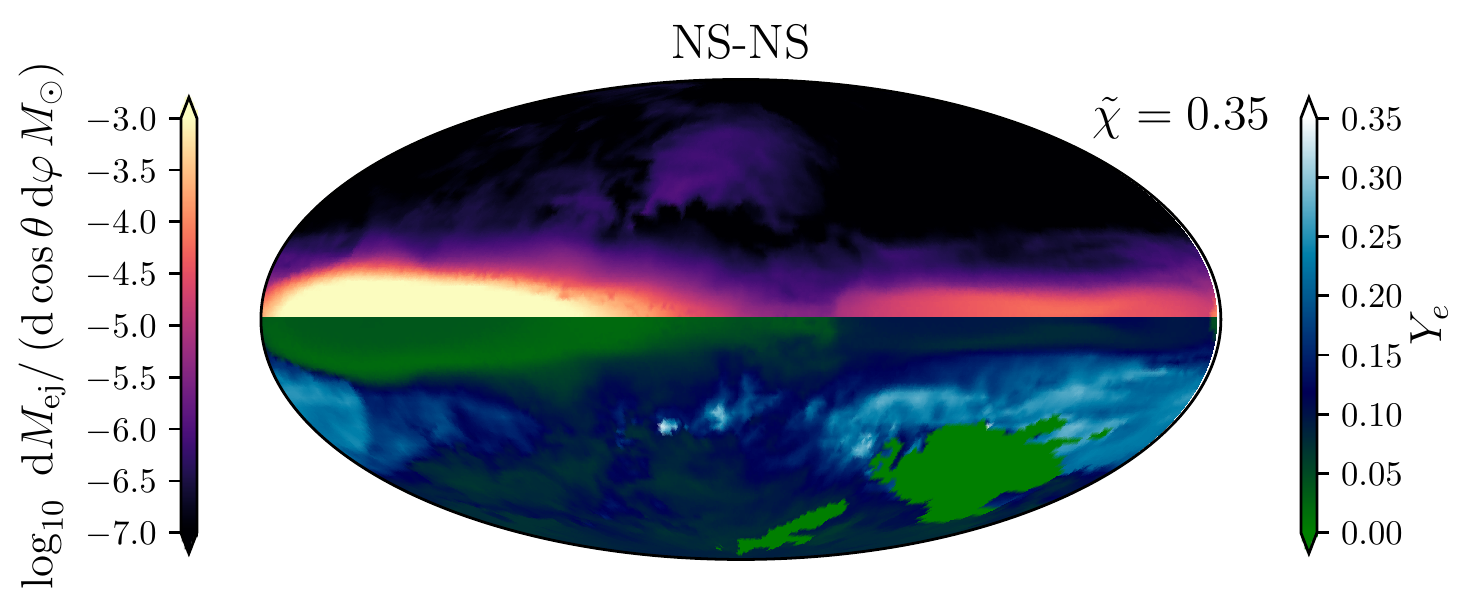}
  \caption{Time-integrated ejected mass $M_{\rm ej}$ and mass-weighted
    average electron fraction $Y_e$ projected onto a sphere at radius
    $r=295\,\rm km$ from the origin. Shown are the results for NS--NS
    (left column) and BH--NS mergers (right column) at a time shortly
    after the two arms have passed the detector. Only unbound fluid
    elements have been recorded. The rows correspond different effective
    spins $\tilde{\chi}$. All models have been computed using the
    \texttt{TNTYST} EOS. We point out that especially at high latitudes
    where almost no mass is ejected, spurious low density dynamics might
    lead to artificially high or low $Y_e$ values in those regions. These
    can safely be ignored as they do not contribute to the overall
    ejected mass.  }
  \label{fig:Mej_2D}
\end{center}
\end{figure*}

The overall dynamical picture described above is modified only slightly by
the presence of spin in the primary star/BH. An non-zero initial spin 
will increase the total angular momentum of the merged object
and, hence, increase the dynamical ejection of matter. This can be seen
in Fig. \ref{fig:XYB}, where we compare the tidal arms in the case of
zero spin and low spin, $\tilde{\chi} = 0.15$. We can see that the fast
low density arm is, indeed, less massive for low spin cases. In order to
assess the dependence on the EOS, the right panels of Fig. \ref{fig:XYB}
show the same simulations performed with the \texttt{BHB}$\Lambda\Phi$
EOS. The \texttt{BHB}$\Lambda\Phi$ EOS has a
slightly smaller maximum mass but a larger tidal deformability.  
As a result, the secondary NSs deform more strongly
under the tidal field of the primary, leading to an enhanced mass
ejection and a larger remnant disk mass in the case of BH--NS merger
\citep{Kyutoku2015,Foucart2018b}. On a more qualitative level, we find an
increase in mass in the second (fast) arm for the zero spin case (compare
left columns of the left and right panels of Fig. \ref{fig:XYB}). In all
of the above cases, the main evolution of the tidal tail seems to be
almost the same as in the BH--NS case. A detailed analysis and comparison
of the properties of the mass ejection will be given in the following.

\subsection{Dynamical mass ejection: geometrical and physical distribution}

\begin{figure*}[t]
  \centering \includegraphics[width=0.9\textwidth]{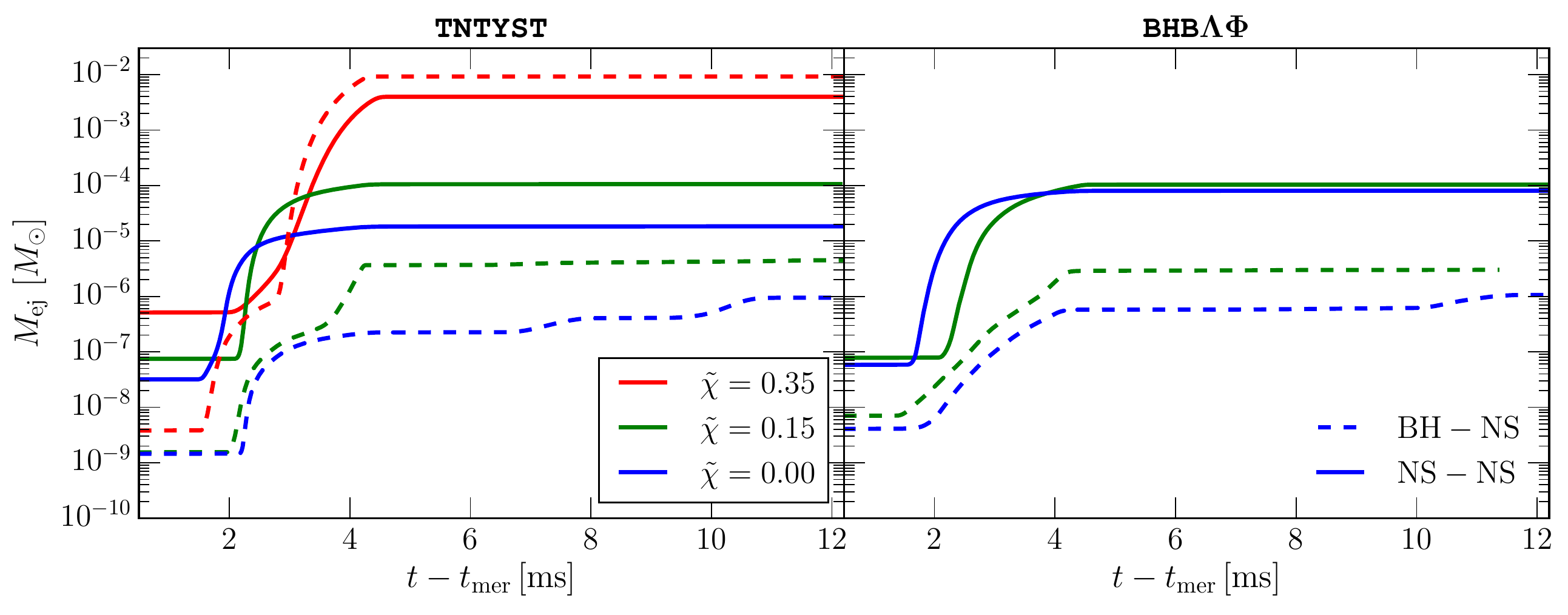}
  \caption{Ejected mass $M_{\rm ej}$ for the different configurations
  as a function of effective spin $\tilde{\chi}$ relative to the time of
  merger $t_{\rm mer}$.}
  \label{fig:Mej}
\end{figure*}

In order to gauge the observational impact of the fast ejecta, we perform a
detailed analysis of its properties and composition.   We record the mass
ejection until the primary (tidal) arm has fully passed the detector (about
$5-10\, \rm ms$ after merger, see also Fig. \ref{fig:Mej}).  In Fig.
\ref{fig:Mej_2D}, we start by comparing the final time-integrated angular
distributions of the ejecta for the \texttt{TNTYST} EOS simulations when
using a Mollweide projection on a spherical surface at a distance of $
295\, \rm km$ from the origin.  Since the secondary arm is in all cases
moving at faster or comparable speeds, it is always fully captured.  We
only record fluid elements that are unbound according to the Bernoulli
criterion, \ie $h u_t < -1$, where $u^\mu$ is the fluid four-velocity and
$h$ the specific fluid enthalpy \citep{Bovard2016}.

\begin{figure*}
\begin{center}
  \centering
  \includegraphics[width=0.35\textwidth, trim=0 0 60  0, clip]{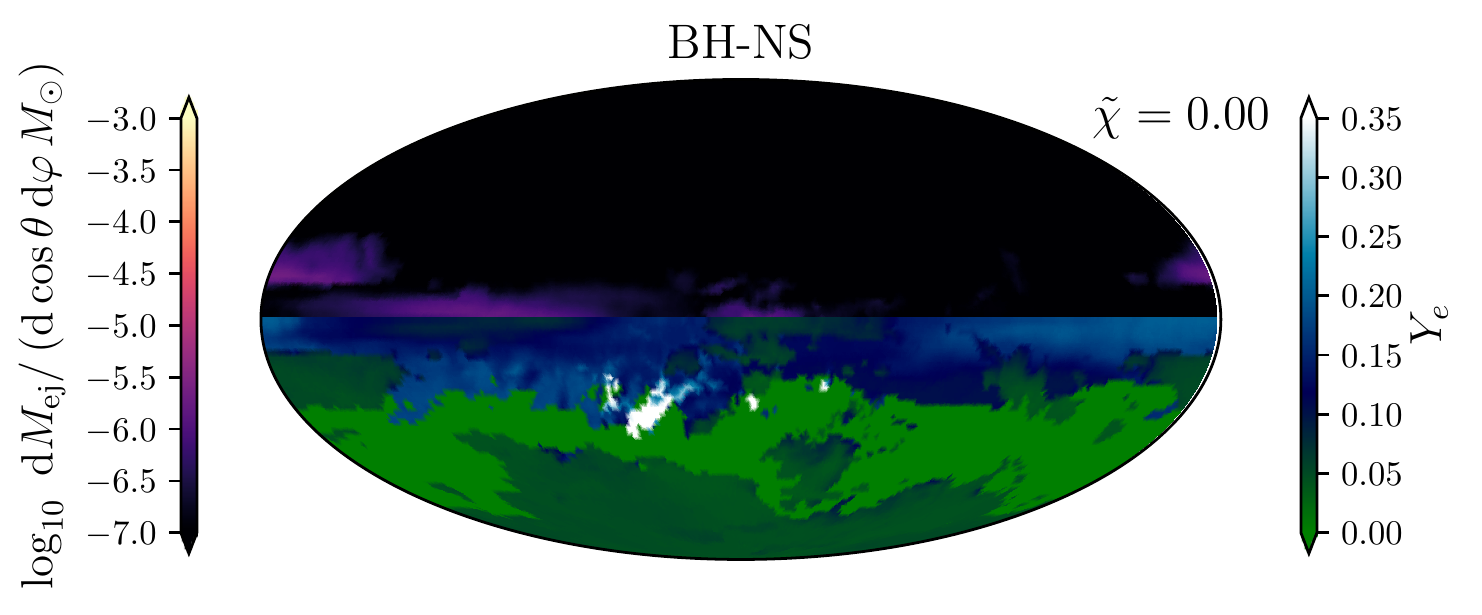}
  \includegraphics[width=0.35\textwidth, trim=60 0 0  0, clip]{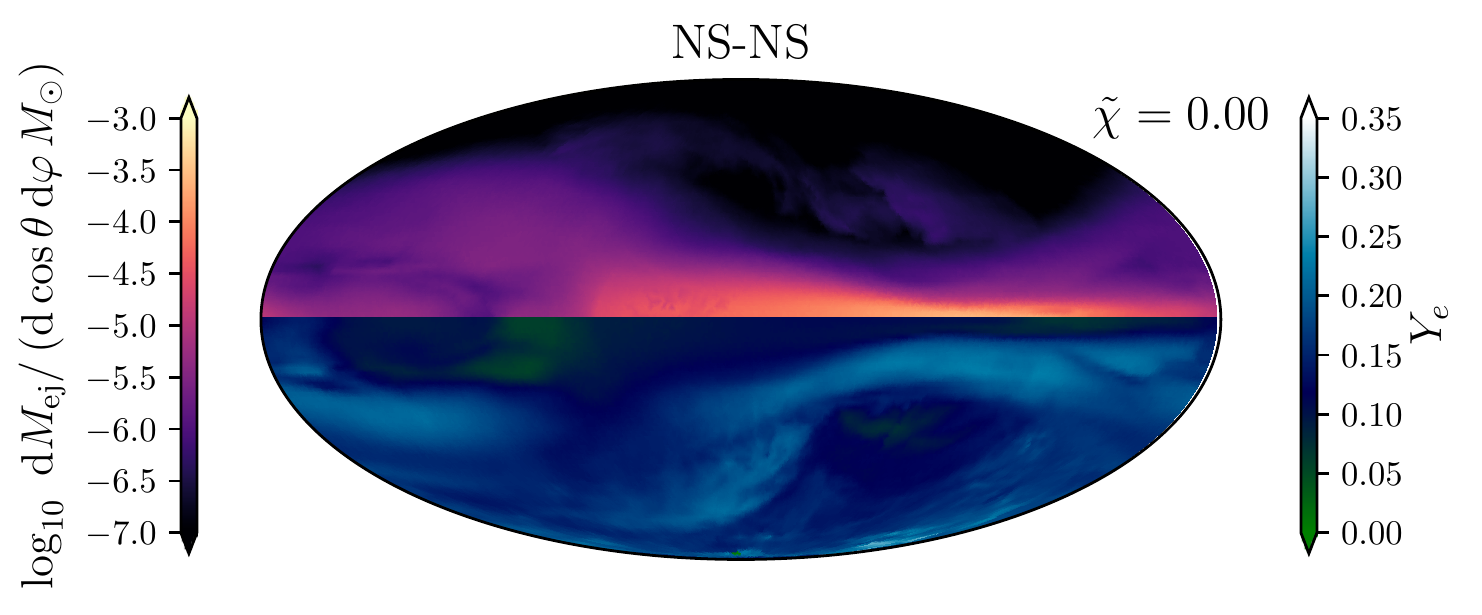}
  \\
  \includegraphics[width=0.35\textwidth, trim=0 0 60 19, clip]{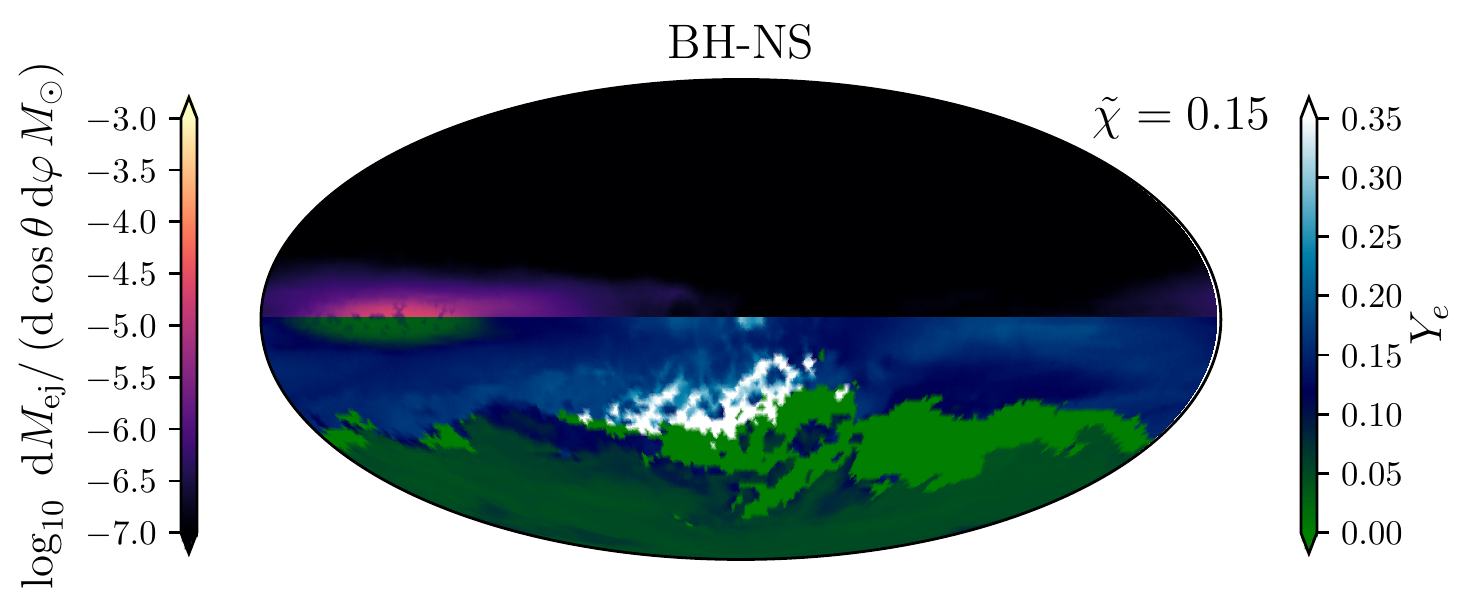}
  \includegraphics[width=0.35\textwidth, trim=60 0 0 19, clip]{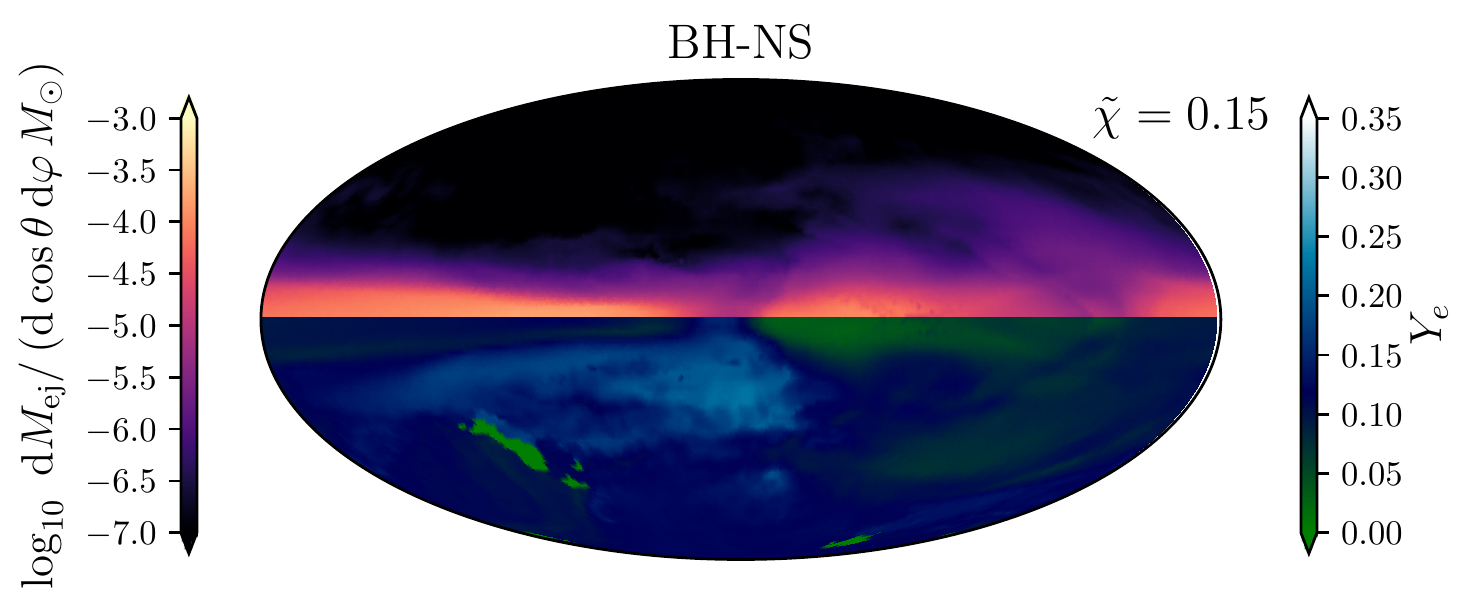}
  \caption{Same as Fig. \ref{fig:Mej_2D} but for the models using the
    \texttt{BHB$\Lambda\Phi$} EOS.}
  \label{fig:Mej_2D_a}
\end{center}
\end{figure*}

Starting with the zero spin case, $\tilde{\chi} =0$ (top row), we note
that for the BH--NS binary (left column) only very small amounts of
ejecta are detected (upper hemispheres) and that the main tidal tail is
absent since the system is too bound to tidally eject mass. For the
NS--NS binary (right column), the situation is very similar in that the
main tidal tail of ejecta is almost absent. However, we also note that the
low-mass fast tail is clearly visible as unbound material with a
significant proton enhancement, \ie with $ Y_e > 0.1$ (lower
hemispheres). While we find that the NS--NS merger can reach temperatures
$>100\, \rm MeV$ at the interface between the two stars, the secondary
arm is most likely sourced by cold material of the primary NS. Hence, the
increased electron fraction is not the result of weak interactions, but
corresponds to densities in the outer layers of the primary star that
correspond to densities with electron fractions $Y_e>0.1$ (in
beta-equilibrium) \citep{Shen2002}. When increasing the effective spin,
$\tilde{\chi} =0.15$ (middle row), we note that the massive tidal tail
becomes unbound and shows up for both BH--NS and NS--NS systems as
neutron-rich, \ie with $Y_e< 0.05$, material ejected
mostly along the equatorial plane. In addition to the tidal tail, also
the fast ejecta are clearly visible (and dominant) for the NS--NS system
(\cf eastern part of the projection). As for $\tilde{\chi} =0$, this fast
tail does come at higher electron fractions and can be clearly
distinguished from the tidal tail, which is ejected in the opposite
direction (\cf western part of the projection). Finally, with even higher
spin $\tilde{\chi} =0.35$, the tidal ejection is significantly enhanced
as expected from previous studies \citep{Foucart2014,Kyutoku2015}. The
secondary arm is now subdominant but still clearly visible in the NS--NS
case.

We contrast these findings with the same results obtained for the stiffer
\texttt{BHB}$\Lambda\Phi$ EOS, which are shown in Fig.
\ref{fig:Mej_2D_a}. While the overall evolution is very similar, for this
EOS the mass ejection is enhanced in the case of zero spin in the NS--NS
case (top right column). Again, we can clearly distinguish between the
two arms by means of the nuclear composition, with the secondary arm
having higher electron fractions. Different from the BH--NS case, we find
that moderate spins, $\tilde{\chi}=0.15$, give rise to an increased tidal
mass ejection (see the bottom row), although the secondary arm is still
the dominant contribution. Overall, the binaries considered here seem to
indicate that the amount of tidal ejecta differs between NS--NS and
BH--NS binaries having the same mass. We thus conjecture that this
difference will mainly depend on the size of the NSs, although follow-up
work will be needed to clarify this point. This difference is already
present in the apparent sizes of the tidal arms as shown in the right
panels of Fig. \ref{fig:XYB}, which are less similar than for the runs
with the \texttt{TNTYST} EOS, see Fig. \ref{fig:XYB}.

\begin{figure*}[t]
  \centering
  \includegraphics[width=0.9\textwidth]{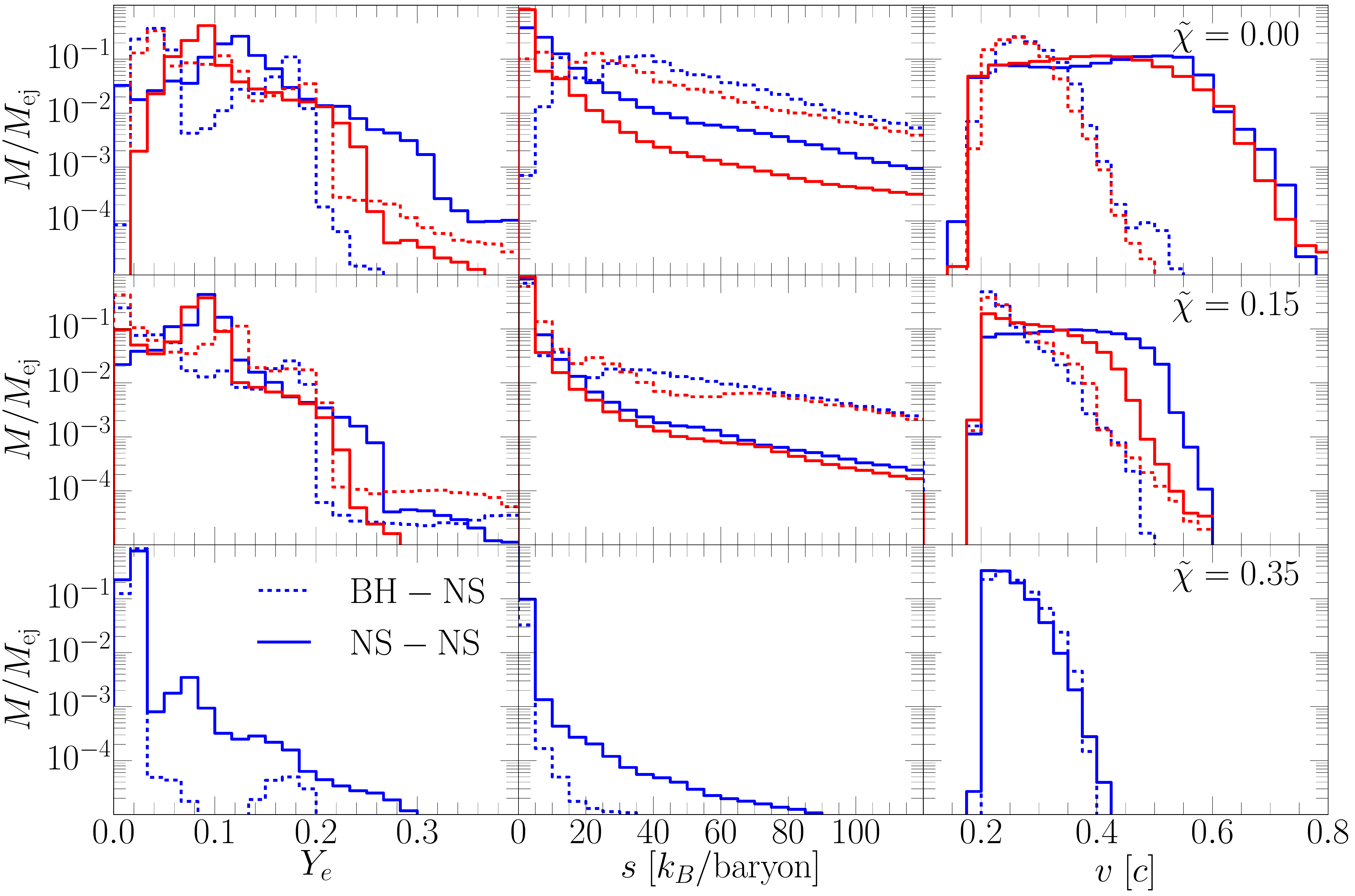}
  \caption{Compositional properties of the dynamical mass ejection. From
    left to right we show the electron fraction $Y_e$, the entropy per
    baryon $s$ and velocity $v$ as computed from the Lorentz factor.
    Solid lines denote NS--NS and dashed lines BH--NS mergers. The rows
    correspond to different effective spins $\tilde{\chi}$. Results are
    shown for the \texttt{TNTYST} (blue) and \texttt{BHB}$\Lambda\Phi$
    (red) EOSs.}
  \label{fig:prop}
\end{figure*}

\begin{figure*}[t]
  \centering \includegraphics[width=0.9\textwidth]{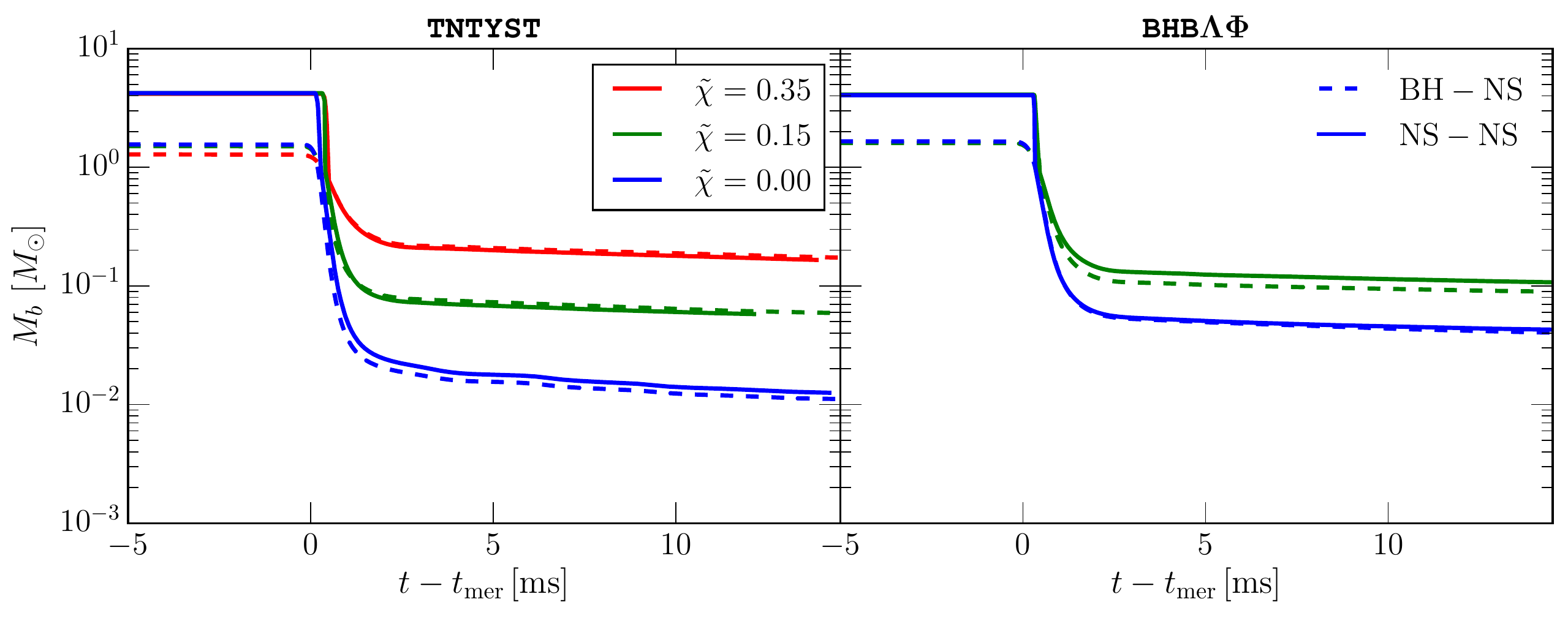}
  \caption{Baryon mass $M_b$ outside of the apparent horizon at any point
    during the simulation. After the sudden drop at merger, these curves
    represent the matter inside the accretion disk (outside of the
    apparent horizon) and in the ejecta.  NS--NS mergers are denoted by
    solid lines, while BH--NS mergers with dashed lines. }
  \label{fig:MB}
\end{figure*}

To complement the information provided so far on the fast tails, we
report in Fig. \ref{fig:prop} the distribution of the relative mass
fraction in terms of the nuclear composition via the electron fraction
$Y_e$ (left columns), of the entropy per baryon $s$ (middle columns), and
of the ejecta velocity $v$ (right columns). Different rows refer to
different spins, lines of different type but of the same color are
representative of either BH--NS (dashed lines) or NS--NS systems (solid
lines), while lines of different color but of the same type are
representative of binaries evolved with the \texttt{TNTYST} (blue lines)
of the \texttt{BHB}$\Lambda\phi$ (red lines) EOS.  As anticipated from
the angular distribution in Fig. \ref{fig:Mej_2D}, we find that NS--NS
systems produce distributions of the electron fraction $Y_e$ that are
shifted towards higher values in the low-spin cases. This can be mainly
attributed to the appearance of the fast ejecta for NS--NS mergers that
dominate the mass ejection. These ejecta come from the outer layers of
the primary star, which have higher $Y_e$ values than in the core in
beta-equilibrium. More specifically, for simulations with the
\texttt{TNTYST} EOS (blue lines), the $Y_e$ distributions peak around
$Y_e\simeq 0.15 \left( 0.10 \right)$ for spins $\tilde{\chi} = 0.00\,
(0.15)$, with a high $Y_e$ tail in the distribution extending up to
$Y_e^{\rm max} \approx 0.3\, (0.35)$. In the high-spin case,
$\tilde{\chi} = 0.35$, for which we have only simulations with the
\texttt{TNTYST} EOS (bottom row), we can also see the appearance of the
fast tail in large-$Y_e$ part of the distribution; however, since this
ejection is overall subdominant, the bulk peaks of the distributions at
$Y_e\lesssim 0.02$ do not shift significantly, as expected for tidal
ejecta that are mainly dominated by highly neutron-rich material from the
NS core. In this case, the distribution of the fast tail peaks around
$Y_e\simeq 0.08$ and extends up to $Y_e < Y_e^{\rm max} \approx 0.18$. As
expected for these type of ejecta they come at very low specific
entropies $s < 20\, k_B/{\rm baryon}$, with the NS--NS distributions
having average entropies that are slightly larger than the corresponding
ones from BH--NS binaries, as it is obvious given that and actual
collision takes place in the first case (see middle column of
Fig. \ref{fig:prop}). Furthermore, the average entropies tend to decrease
as the spin in the system is increased, underlining that the disruption
process is less energetic with more rapidly spinning primaries.

Crucially, the existence of a secondary tidal arm in the NS--NS systems
gives rise to a velocity distribution that has a large and massive portion
with velocities $0.5 \lesssim v \lesssim 0.8$. This can be clearly seen
in the top-right panel of Fig. \ref{fig:prop}. Note also that because
these high-velocity tails are absent in the corresponding BH--NS mergers,
they certainly represent the most important difference discriminating
BH--NS and NS--NS systems having the same mass. As the spin of the
primary is increased, the high-velocity become less pronounced.
In fact, the fast-ejecta tails essentially disappear in the $\tilde{\chi}=0.35$
case, where BH--NS and NS--NS mergers yield overall the same velocity
distributions (see bottom right, Fig. \ref{fig:prop}). Although there is
still considerable mass in this secondary tidal arm (see bottom left,
Fig.  \ref{fig:Mej_2D}), it has the same velocities as the main tidal
tail ejecta. This behaviour is easy to interpret: in the
$\tilde{\chi}=0.35$ case, the primary star is already spinning rapidly
and a further sudden spin-up through the merger process does not have
lead to the ejection of rapidly moving matter.

So far, the previous discussion only weakly depends on the EOS and,
overall, evolutions with the \texttt{BHB}$\Lambda\Phi$ EOS (red lines)
give very similar distributions, with the minor difference that the
distributions cut-off earlier at high $Y_e$ and for the zero-spin case
are shifted to slightly lower electron fractions, \ie $Y_e \simeq 0.1$.

\centerwidetable
\begin{deluxetable*}{|l|c|c|c|c|c|c|c|}
	\tablewidth{0pt} \tablecaption{Properties of the ejected and
          bound mass after the merger.  Shown are the values of the
          ejected baryon mass for BH--NS and for NS--NS mergers.  In the
          latter case, we also give the mass of the ejecta having speeds
          $v\,>\,0.4$, $M_{\rm ej}^{\rm fast},$ and the kinetic energy,
          $T_{\rm ej}$. We finally report the remnant baryon mass,
          $M_b^{\rm fin}$, that has not been accreted by the end of the
          simulation and contrast it with the analytic estimate $M_b^{\rm
            FHN}$ of \citet{Foucart2018b}.
    \label{tab:Mej}}
	
	\tablehead{
    & $M_{\rm ej}\,\left[M_\odot\right]$  & $M_{\rm ej}^{\rm fast}\,\left[M_\odot\right]$ 
    &$T_{\rm ej}\, \rm [erg]$ & $M_b^{\rm fin}\,\left[M_\odot\right]$ & $M_b^{\rm FHN}\,\left[M_\odot\right]$ 
    & $\tilde{\chi}$
	}

	\startdata
    $\texttt{TNT.BH.chit.0.00}$ 		& $9.5\times 10^{-7}$	& $1.6\times 10^{-9}$  	
    		& $4.3\times 10^{48}$ 		& $1.3\times 10^{-2}$	& $6.4\times 10^{-3}$	& $0.00$	\\
    $\texttt{TNT.BH.chit.0.15}$	 	& $4.5\times 10^{-6}$	& $1.1\times 10^{-8}$
    		& $1.6\times 10^{49}$ 		& $5.8\times 10^{-2}$	& $5.1\times 10^{-2}$	& $0.15$	\\
    $\texttt{TNT.BH.chit.0.35}$		& $9.2\times 10^{-3}$ 	& $1.4\times 10^{-8}$	
    		& $3.0\times 10^{50}$ 		& $1.7\times 10^{-1}$ 	& $1.8\times 10^{-1}$ 	& $0.35$	\\
	\tableline                                                                                
    $\texttt{TNT.NS.chit.0.00}$		& $1.8\times 10^{-5}$ 	& $1.1\times 10^{-5}$ 
    		& $1.0\times 10^{47}$		& $1.1\times 10^{-2}$ 	& -- & $0.00$ \\
    $\texttt{TNT.NS.chit.0.15}$ 		& $1.1\times 10^{-4}$  	& $3.2\times 10^{-5}$ 
    		& $3.6\times 10^{47}$ 		& $5.4\times 10^{-2}$ 	& -- & $0.15$ \\
    $\texttt{TNT.NS.chit.0.35}$ 		& $4.0\times 10^{-3}$ 	& $3.4\times 10^{-7}$ 
    		& $7.6\times 10^{50}$ 		& $1.5\times 10^{-1}$ 	& -- & $0.35$ \\ 
   	\tableline
	\tableline                                                                                               
    $\texttt{BHBLP.BH.chit.0.00}$ 	& $1.1\times 10^{-6}$  	& $1.1\times 10^{-9}$ 
    		& $1.0\times 10^{47}$ 		& $3.3\times 10^{-2}$ 	& $3.6\times 10^{-2}$ 	& $0.00$ \\
    $\texttt{BHBLP.BH.chit.0.15}$ 	& $3.0\times 10^{-6}$  	& $8.7\times 10^{-9}$ 
    		& $2.5\times 10^{47}$ 		& $7.3\times 10^{-2}$	& $1.0\times 10^{-1}$ 	& $0.15$ \\
	\tableline                                                                                  
    $\texttt{BHBLP.NS.chit.0.00}$ 	& $7.9\times 10^{-5}$ 	& $3.9\times 10^{-5}$ 
    		& $2.8\times 10^{48}$ 		& $3.5\times 10^{-2}$ 	& -- & $0.00$ \\
    $\texttt{BHBLP.NS.chit.0.15}$ 	& $1.0\times 10^{-4}$ 	& $6.3\times 10^{-6}$ 
    		& $1.1\times 10^{49}$ 		& $8.7\times 10^{-2}$	& -- & $0.15$ \\
	\enddata

\end{deluxetable*}

%
It should be remarked that the fast ejecta are produced over a very short
window in time around the merger. This is clearly summarized in
Fig. \ref{fig:Mej}, which reports the evolution of ejected rest-mass
$M_{\rm ej}$ for the different configurations as a function of effective
spin $\tilde{\chi}$ relative to the time of merger $t_{\rm mer}$. Note
that the spin in the primary is responsible for a considerable
amplification of the amount of ejected mass (\cf the $\tilde{\chi}=0$ and
$\tilde{\chi}=0.35$ cases). While this is more sever for BH--NS systems --
where the amplification is of about four orders of magnitude -- the
ejected mass can grow by a factor $\lesssim 100$ in the case of NS--NS
systems.

Furthermore, because of the rapidity with which mass is ejected, the
amount of mass that is present in the disk around the newly formed BH
will not differ significantly between the BH--NS and the NS--NS binaries
considered here. This is shown in Fig. \ref{fig:MB}, which reports the
baryon mass $M_b$ outside of the apparent horizon at any point during the
simulation; here too, different lines refer to either BH--NS mergers
(dashed lines) or to NS--NS mergers (solid lines). After the sudden drop
at merger, these curves essentially represent the combined matter inside
the accretion disk as well as in the ejecta.  By comparing the remnant
baryon masses after the merger for the various binaries it is apparent
that the disks formed in both mergers are essentially the same and that
the fast ejecta constitute only a small fraction of the total baryonic
mass outside the BH. We summarize this in Table \ref{tab:Mej}, which
collects the properties of the ejected and bound mass after the merger
for the various binaries considered. Remarkably, all mergers with
different spins and different EOSs yield comparable amounts of baryon
mass for BH--NS and NS--NS systems, which are in good agreement with
previous estimates \citep{Foucart2018b}.

As a concluding remark, we discuss the prospects that these fast ejecta
may actually be detectable and leave an imprint that is different from
the standard afterglow emission coming from the disk winds of the remnant
accretion disk \citep{Fernandez2015, Fernandez2015b, Siegel2017}. Since
both systems start out with the same gravitational masses and spins, the
final spacetime after merger, \ie the properties of the final BH, will be
very similar in both cases. As the main differences in the disk ejection
can only be driven by different properties of the accretion disks and
these do not vary significantly across the binaries considered (\cf Fig.
\ref{fig:MB}), the afterglow emission coming from the disk winds should
be comparable in both cases. Hence, the most visible impact that the fast
ejecta may potentially leave is in the kilonova emission, whose modelling
needs to take into account the (asymptotic) ejecta velocity achieved in
the homologous-expansion phase and which obviously depends on the
velocity at the ejecting site \citep[see discussion
  in][]{Bovard2017}. Indeed, the fast ejecta launched at merger could
lead to a bright kilonova precursors just hours after the merger
\citep{Metzger2015}, or to a synchrotron emission several years later
\citep{Hotokezaka2018}. We will comment on this further in the next
section.

\section{Discussion}
\label{sec:discussion}

We have explored and contrasted the dynamics of binary NS--NS and
BH--NS mergers that could be representative of high-mass
gravitational-wave events such as GW190425 \citep{Abbott2020}. By
choosing total masses and mass ratios such that the primary mass $m_1$ is
close to the maximum mass of nonrotating NSs, we were able to investigate
a region of the space of parameters where a realistic overlap between the
two scenarios exists. By performing fully general-relativistic
magnetohydrodynamics simulations of the merger of the two types of binary
systems, we were able to show that the main mass ejection mechanism
through the secondary, as well as the remnant disk mass is comparable in
both cases.

While this indicates that the kilonova emission expected for these
systems \citep{Fernandez2015,Fernandez2015b,Siegel2017}, is likely going
to be very similar, we identified another mass-ejection channel that
could help to distinguish a primary NS from a primary BH in the two
binary systems. More specifically, we found that in a NS--NS system the
presence of a massive NS and the consequent spin-up at merger through the
torques exerted by the secondary leads to a second burst of mass ejection
propagating in the direction opposite to that of the main tidal tail.  By
carefully analysing the matter properties in terms of its geometrical and
physical distributions, we were able to show that the second
mass-ejection burst is propagating at very high speeds, $0.5 \lesssim v
\lesssim 0.8$ for irrotational binaries. This is considerably larger than
for a BH--NS system with the same mass, whose velocity distribution is $v
\lesssim 0.5$, when comparing the second tidal arm produced by the
spin-up of the primary star with the first tidal arm produced by the
disruption of the secondary star, we have shown that the former is also
considerably more proton rich, with a peak in the distributions at
$Y_e\approx 0.12$. Hence, these high-velocity tidal certainly represent
the most important difference discriminating BH--NS and NS--NS systems
having the same (high) mass.

It has been proposed that such fast dynamical ejecta produced at merger
can lead to bright kilonova precursors just hours after the merger
\citep{Metzger2015}, or to a synchrotron emission several years later
\citep{Hotokezaka2018}. However, so far searches for this X-ray
re-brightening in the afterglow of GW170817 have remained inconclusive
\citep{Troja2020}. In particular, in the case of a precursor to the
kilonova afterglow, \citet{Metzger2015} found that for optimistic
estimates of the opacity, an amount of ejecta of $\sim 3\times 10^{-5}\,
M_\odot$, are sufficient to produce a beta-decay powered reheating of the
afterglow, which would lead to a luminosity peak at $\sim 1$ hour after
the merger. While this matches well with our results, we caution that
because the fast ejecta we have found in the NS--NS systems are less
neutron rich, this will result in a reduced availability of free neutrons
to undergo beta-decay. Additionally, the strong concentration of these
ejecta along the equatorial plane would also reduce the detectability
prospects, especially for a transient that would only be visible in the
first few hours after merger \citep{Metzger2015}. On the other hand, if
the interaction of the fast ejecta with the interstellar medium would
indeed produce a re-brightening of the afterglow \citep{Hotokezaka2018},
we estimate that even in the most optimistic case, \ie
\texttt{BHBL.chit.0.15}, a radio peak flux of $F_{\nu} \lesssim 1\,
\mu\rm Jy$ (at 3 GHz), which would likely not be detectable \citep[see
  Sec. 5.1 of][for details]{Radice2018a}.

Besides the present objective difficulties in observing the presence of
these fast ejecta, we should also comment on the limitations of our
numerical study. By its very nature, tidal disruption crucially depends 
on the radius of the NS. It has been shown
that large stars, which are affected more strongly by tidal forces,
feature enhanced mass ejection
\citep{Foucart2013a,Kyutoku2015,Foucart2018b}. To address this
we have chosen only two EOSs spanning the range of radii compatible with GW170817
\citep{Most2018}, without fully exploring it.
These two EOS naturally come with fixed maximum masses that are not far
apart. In particular, should the maximum mass be much higher
than $M_{_{\rm TOV}} \gg 2.3\, M_\odot$ \citep{Tan2020,Fattoyev2020}, a
regime of even higher mass ratios should also be explored.

Finally, for obvious computational costs, we have limited ourselves to
only a single total mass, $M_{\rm tot} = 3.6\, M_\odot$, restricting the
mass ratio, the mass and spin of the secondary. Future work will be
needed to elucidate this dependence, in particular with respect to the
fast tail of the mass ejection for NS--NS mergers. Such a study will need
to be accompanied by a careful (re-)analysis of lower mass NS--NS
systems, in order to understand the maximum mass ratio at which fast
ejecta along the equatorial plane are to be expected
\citep{Bernuzzi2020}.

\section*{Acknowledgements}

ERM thanks Carolyn Raithel for insightful discussions on neutron-star
composition. ERM gratefully acknowledges support from a joint fellowship
at the Princeton Center for Theoretical Science, the Princeton Gravity
Initiative and the Institute for Advanced Study. The simulations were
performed on the national supercomputer HPE Apollo Hawk at the High
Performance Computing Center Stuttgart (HLRS) under the grant number
BBHDISKS. The authors gratefully acknowledge the Gauss Centre for
Supercomputing e.V. (www.gauss-centre.eu) for funding this project by
providing computing time on the GCS Supercomputer SuperMUC at Leibniz
Supercomputing Centre (www.lrz.de). LR gratefully acknowledges funding
from HGS-HIRe for FAIR; the LOEWE-Program in HIC for FAIR; ``PHAROS'',
COST Action CA16214.


\software{Einstein Toolkit \citep{loeffler_2011_et},
          \texttt{Carpet} \citep{Schnetter-etal-03b},
          \texttt{AHFinderDirect} \citep{Thornburg2003:AH-finding}, 
          \texttt{Frankfurt-/IllinoisGRMHD (FIL)} \citep{Most2019b,
	  Etienne2015},
	  \texttt{LORENE} (\url{https://lorene.obspm.fr}), 
	  \texttt{Kadath}\citep{Grandclement09}
          }
          
\bibliographystyle{yahapj}
\bibliography{aeireferences}

\end{document}